# Enhancement of bulk second-harmonic generation from silicon nitride films by material composition

K. Koskinen,[1,*] R. Czaplicki,[1] A. Slablab,[1] T. Ning,[2] A. Hermans,[3,4] B. Kuyken,[3,4] V. Mittal,[5] G. S. Murugan,[5] T. Niemi,[1] R. Baets,[3,4] and M. Kauranen[1]

[1]Laboratory of Photonics, Tampere University of Technology, 33101 Tampere, Finland
[2]School of Physics and Electronics, Shandong Normal University, Jinan, Shandong 250014, China
[3]Photonics Research Group, INTEC Department, Ghent University-imec, Ghent, Belgium
[4]Center for Nano- and Biophotonics, Ghent University, Technologiepark-Zwijnaarde 15, 9052 Ghent, Belgium
[5]Optoelectronics Research Center, University of Southampton, Southampton SO17 1BJ, United Kingdom
*Corresponding author: kalle.o.koskinen@tut.fi

We present a comprehensive tensorial characterization of second-harmonic generation from silicon nitride films with varying composition. The samples were fabricated using plasma-enhanced chemical vapor deposition, and the material composition was varied by the reactive gas mixture in the process. We found a six-fold enhancement between the lowest and highest second-order susceptibility, with the highest value of approximately 5 pm/V from the most silicon-rich sample. Moreover, the optical losses were found to be sufficiently small (below 6 dB/cm) for applications. The tensorial results show that all samples retain in-plane isotropy independent of silicon content, highlighting the controllability of the fabrication process.

*OCIS codes: (190.0190) Nonlinear optics; (190.4400) Nonlinear optics, materials; (310.6860) Thin films, optical properties.*

High-performance complementary metal oxide semiconductor (CMOS) compatible materials are essential elements for advanced on-chip photonic devices to realize the future progress in all-optical processing. The ultra-fast speed and high bandwidth of integrated photonic networks continuously require new materials possessing excellent linear and nonlinear optical properties [1, 2]. Although silicon (Si) is still the most commonly used CMOS material, the intrinsic drawbacks of Si, such as its narrow bandgap and centrosymmetric structure highly limit its future applications especially in the visible and ultraviolet spectral regimes [2, 3]. Thus, exploring novel CMOS-compatible materials with wide bandgap and strong optical nonlinearities is very important for future integrated devices.

Many photonic applications rely on nonlinear optical effects. One of the limitations of many nonlinear materials for CMOS-compatible platforms is the lack of second-order nonlinearity due to centrosymmetry. The problem can be overcome by poling [4, 5], straining the material [3] or by using multilayer composites [6-8]. Unexpectedly, CMOS-compatible amorphous silicon nitride films (SiN) have been shown to possess a bulk second-order nonlinearity by measuring strong second-harmonic generation (SHG) from thin films [9-11]. Although the exact reason for this strong SHG response remains unclear, it is believed that the complicated composition, crystalline phase and defects in the film during the deposition may be responsible [10, 12-16].

In this Letter, we show that the strong second-harmonic signal from SiN films can be further enhanced by varying the composition of the films prepared with plasma-enhanced chemical vapor deposition (PECVD). Furthermore, we demonstrate that such composition tuning does not compromise the linear optical properties or optical losses of the material for applications. Our results are crucial for the comprehensive understanding of the linear and nonlinear optical properties in SiN films with different structures, opening the path for further optimization of SiN for on-chip devices.

We recognize that there have been previous studies yielding different values for the SHG susceptibility of SiN [9, 10, 11, 17, 18]. Samples prepared by sputtering can yield very high values of the susceptibility. Unfortunately, the susceptibility value depends extremely sensitively on material composition [9] or the samples possess varying symmetry [11]. These results suggest that the sputtering process can be poorly controlled. In contrast, our PECVD process is consistent, maintaining sample isotropy about the surface normal [10, 17, 19]. PECVD is also compatible with the thermal budget of finished CMOS-circuits. Sample composition is also an important parameter for electric-field-enhanced SHG [18], but this preliminary study reported only a scalar value for the susceptibility. Our tensorial results, obtained through a very advanced model, combined with loss measurements are thus crucial in addressing the suitability of SiN in various photonic applications.

SiN films of thicknesses of approximately 500 nm and of different compositions were grown on fused silica substrates using the PECVD technique. Four samples (S10, S20, S30 and S40) were fabricated in the Laboratory of Photonics (Tampere University of Technology) with the reactive gas mixture of 2% $SiH_4/N_2$ and $NH_3$, process pressure of 1000 mTorr, and deposition temperature of 300°C (Plasmalab 80 plus, Oxford Instruments). For these four samples, the plasma was generated using a RF field with frequency of 13.56 MHz and power of 20 W. The material composition of the samples was controlled by adjusting the flow rate of $NH_3$ (10, 20, 30 and 40 sccm for samples labeled S10, S20, S30 and S40, respectively) while simultaneously applying a constant flow rate of 2% $SiH_4/N_2$ of 1000 sccm. In order to further address the role of the fabrication procedure, we prepared one additional sample (S35, 35 sccm of $NH_3$) with PECVD (Advanced Vacuum Vision 310 PECVD) at Ghent University-imec using a gas mixture of $SiH_4$ (40 sccm), $NH_3$ (35 sccm) and $N_2$ (1960 sccm) under deposition temperature of 270°C and process pressure of 650 mTorr. For the S35 sample, the plasma was generated using an exciting field alternating between one second period of high frequency field (13.73 MHz, 30 W) and five second period of low frequency field (~100-300 kHz, 50 W). We also addressed a SiN thin film fabricated using low pressure chemical vapor deposition (LPCVD). However, the SHG response from the LPCVD sample was found to be extremely weak and it will not be discussed any further here.

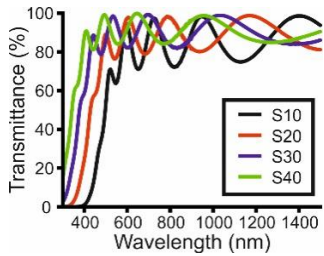

Fig. 1. Normalized transmittance spectra of samples with different compositions.

The fabricated samples were characterized by linear optical spectroscopy (UV-VIS-NIR spectrophotometer, Shimadzu UV-3600) for wavelengths from 300 nm to 1500 nm (Fig. 1). As the silicon content increases (with the lower flow rate of $NH_3$), the transmittance threshold shifts towards shorter wavelengths in good agreement with previous studies [9]. The oscillatory behavior of the transmittance at longer wavelengths can be ascribed to interference between beams reflected at the interfaces of the SiN film. The optical bandgap energies were estimated from a Tauc plot (not shown) to be between ~3 and ~2 eV from the least to the most silicon rich sample.

The wavelength dependent refractive index and thickness of the samples were determined by ellipsometric measurements. The real and imaginary parts of the refractive index are shown in Figs. 2(a) and 2(b), respectively. The refractive indices at the fundamental and second-harmonic wavelengths as well as film thicknesses are shown in Table 1 for all of the studied samples.

Table 1. Thicknesses and refractive indices at fundamental (n) and second-harmonic (N) wavelengths from ellipsometric measurements. The number in the sample name refers to $NH_3$ sccm, as described in the text.

| sample | thickness [nm] | n @ 1064 nm | N @ 532 nm |
|---|---|---|---|
| S10 | 662 | 2.174+0.002i | 2.354+0.022i |
| S20 | 604 | 2.005+0.000i | 2.099+0.007i |
| S30 | 537 | 1.945+0.000i | 1.989+0.002i |
| S35 | 500 | 1.969+0.000i | 2.027+0.002i |
| S40 | 505 | 1.902+0.000i | 1.951+0.001i |

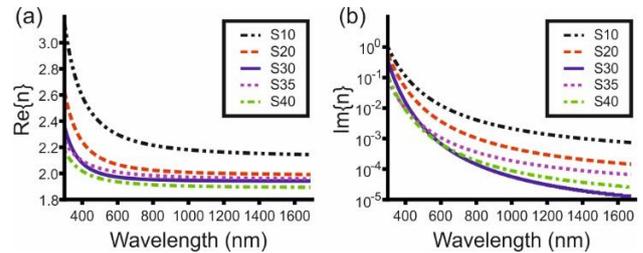

Fig. 2. Real (a) and imaginary (b) parts of refractive indices of the studied SiN thin films with different compositions.

To assess the potential of SiN for applications, we carried out optical loss measurements for all samples at the 633 nm and 1550 nm wavelengths. All samples were found to exhibit losses of less than 6 dB/cm, decreasing further below 3 dB/cm for samples S10 and S20, which is in line with the values previously reported for SiN [20].

The nonlinear measurements were conducted using the setup described in [10]. A mode-locked Nd:YAG laser with a wavelength of 1064 nm, pulse duration of 70 ps, and repetition rate of 1 kHz was used as the source of fundamental light. The spot diameter at the sample was estimated to be a few hundred micrometers. The polarization state of the fundamental beam was controlled with a high-quality polarizer and a motorized quarter-wave plate (QWP). The polarization of the detected SHG signal was selected using another polarizer in front of a photomultiplier tube.

In order to characterize the nonlinear susceptibility, we first illuminated the film at an oblique incidence and studied the polarization signatures of the SHG process for four different polarization controlled measurements. This method is known to uniquely address the relative values of the non-vanishing SHG susceptibility tensor components for samples of $C_{\infty v}$ symmetry [21-23], which is compatible with the known structure for SiN films fabricated using PECVD [10]. Subsequently, we carried out a fifth measurement for fixed polarization states of the incident beam and detected SHG while varying the angle of incidence, followed by a reference measurement from a quartz plate with a known SHG susceptibility to calibrate the value of a single tensor component in absolute units. By combining these two sets of experiments, the values of all nonvanishing SHG susceptibility tensor components can be determined.

For the polarization measurements, we write the SHG field outside the sample as [24]

$$\mathbf{E}_{SHG} = \hat{\mathbf{P}} f^p (e^p)^2 + \hat{\mathbf{P}} g^p (e^s)^2 + \hat{\mathbf{S}} h^s e^p e^s, \qquad (1)$$

where $\hat{\mathbf{P}}$ ($\hat{\mathbf{S}}$) is the unit vector of the polarization components of the SHG field parallel (perpendicular) to the plane of incidence, $e^p$ ($e^s$) is the amplitude of the polarization component of the fundamental field parallel (perpendicular) to the plane of incidence evaluated prior to the sample, and $f^p$, $g^p$, and $h^s$ are auxiliary expansion coefficients describing the polarization signatures of the SHG response.

The expansion coefficients $f^p$, $g^p$, and $h^s$ have been previously shown to be linear combinations of the non-vanishing SHG susceptibility tensor components, which for samples of $C_{\infty v}$ symmetry are $c_{xxz} = c_{xzx} = c_{yyz} = c_{yzy}$, $c_{zxx} = c_{zyy}$, and $c_{zzz}$ [22], where $z$ is the sample normal and $x$, $y$ are the two orthogonal in-plane directions [23]. However, it was recently discovered that in order to properly characterize a film with thickness much smaller than the spot size of the fundamental beam, effects arising from multiple reflections within the films can significantly influence the final results [25]. Thus, we utilize a complete model based on the Green's function formalism for nonlinear optics, which includes all effects arising from reflections [26]. Even in this case, the expansion coefficients can be written as functions of the SHG tensor components as

$$\begin{bmatrix} f^p \\ g^p \\ h^s \end{bmatrix} = M \begin{bmatrix} c_{xxz} \\ c_{zxx} \\ c_{zzz} \end{bmatrix}, \qquad (2)$$

where the matrix $M$ depends only on the experimental geometry and the linear material parameters of the nonlinear film and the substrate. The evaluation of $M$ is a straightforward process but requires arduous calculus, and its full description is omitted due to the extreme length of the mathematical expressions.

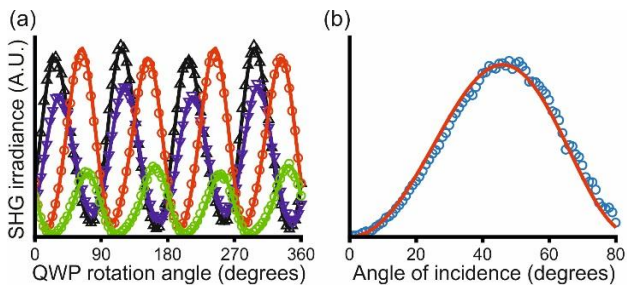

Fig. 3. (a) Experimental data (markers) from polarization controlled SHG (black: $p+s$ in, $p$ out; blue: $p+s$ in, $s$ out; red: $p-s$ in, $p$ out; green: $p-s$ in, $p+s$ out) for sample S30 and analytical fits (lines) to the tensor components. (b) Experimental data of $p$-polarized SHG signal arising from sample S30 with $s$-polarized input versus angle of incidence from the calibration experiment (blue circles) and analytical fit (red line). Similar results were obtained for samples S10, S20, S35 and S40.

To probe the polarization signatures of the SHG process, we chose the four different combinations of polarizations for the detected SHG beam and the initial fundamental beam incident to the QWP to be the same as in [10] and an angle of incidence of 60°. The results from all four measurements were simultaneously fitted for relative values of $c_{xxz}$, $c_{zxx}$ and $c_{zzz}$ and are shown in Fig. 3(a) for the S30 sample. The absolute values of the SHG susceptibility were determined by comparing an angle of incidence controlled calibration measurement of $p$-polarized SHG generated from $s$-polarized input to a measurement from ~1 mm thick Y-cut quartz crystal wedge plate with a known SHG susceptibility of $c_{xxx}^Q = 0.80$ pm/V [(Fig. 3(b)] [27].

The determined values of the SHG susceptibility tensor components of the SiN films of varying composition are shown in Table 2. Our results show that the second-order response can be greatly enhanced by adjusting the flow rate of $NH_3$ during the PECVD fabrication process, showcasing the tunable nature of SiN as a material. Furthermore, the highest bulk-susceptibility of the studied SiN compositions (sample S10) was found to be about 5 pm/V, which is a reasonable value for potential applications. Another interesting result was the discrepancy between the determined susceptibility for S30 sample (for example, $c_{zzz} = 1.10$ pm/V) and the value previously reported for a SiN film prepared under identical conditions ($c_{zzz} = 2.47$ pm/V) [10]. We believe that the difference is due to our present analysis being based on a more advanced model taking reflection effects into account for all experiments, whereas the analysis reported in [10] was based on a simpler approach. This result further highlights the crucial role of a sufficiently detailed model in the nonlinear characterization of thin films.

Note also that the results for the silicon poor samples (S40 and S35) essentially fulfill the Kleinman symmetry $c_{xxz} = c_{zxx}$, as expected for non-resonant nonlinearity, whereas the silicon-richer samples (S30, S20, and S10) start deviating from this symmetry as the resonance for the second-harmonic wavelength is approached (see Fig. 1).

The results also show that the nonlinear response depends sensitively on the fabrication conditions as the sample S35, fabricated at Ghent University, deviates from the general trend of the remaining samples, fabricated at Tampere University of Technology.

Table 2. The calibrated values of second-order susceptibility tensor components of the studied SiN thin films with different compositions.

| sample | $c_{zzz}$ [pm/V] | $c_{xxz}$ [pm/V] | $c_{zxx}$ [pm/V] |
|---|---|---|---|
| S10 | 5.10 | 1.60 | 1.40 |
| S20 | 1.70 | 0.87 | 0.72 |
| S30 | 1.10 | 0.40 | 0.34 |
| S35 | 0.66 | 0.20 | 0.21 |
| S40 | 0.80 | 0.23 | 0.22 |

In conclusion, we have conducted a comprehensive analysis using a detailed analytical model of the SHG response from SiN thin films fabricated using the PECVD method in order to study the dependence of the response on material composition and to determine the optimal composition for nonlinear photonic applications. We have done so by studying five different samples fabricated in two different laboratories with varying fabrication parameters resulting in varying material composition.

Our results show that silicon nitride can be optimized for efficient bulk-type second-harmonic generation response through the material composition at least by a factor of 6, with the value of the highest susceptibility component of approximately 5 pm/V corresponding to the most silicon-rich sample. Furthermore, the optical losses were found to be sufficiently low for viable applications. We believe that our results open the path towards using SiN in a variety of new nonlinear optical applications.

**Funding.** Academy of Finland (265682). K.K. acknowledges the Vilho, Yrjö and Kalle Väisälä Foundation for a personal fellowship. A.H. acknowledges the Research Foundation – Flanders (FWO) for personal fellowship.